\documentstyle[preprint,prb,aps]{revtex}
\begin{document}
\draft
\title{Tunneling into a periodically modulated  
Luttinger liquid.}
\author{A. Gramada \cite{pos} and M. E. Raikh}
\address{Department of Physics, University of Utah, Salt Lake City, Utah  84112}

\maketitle

\begin{abstract}
   We study the tunneling into the edge of the Luttinger liquid
 with periodically
modulated concentration of electrons. It is shown that the modulation,
by causing a gap in the spectrum of plasmons, 
leads to an additional anomaly in the density of states
 at frequency corresponding to the center of the gap. 
The shape of the anomaly depends strongly on the {\em phase} of the
modulation. The sensitivity to the phase is related to the
   plasmon mode, localized at the edge, its  frequency
lying  within the gap
(analog of the Tamm state for an electron in a periodic potential)   
\end{abstract}
\section{Introduction}
\label{sec:I}

The explicit solution of the model of one-dimensional interacting
electron gas \cite{Mat} (Luttinger liquid)
allowed to describe analytically the gap in the tunnel
density of states for the region of energies close to the
Fermi level. The physical origin of the gap is that, in order to add  an
electron to the interacting system, a shift, vacating an extra space,
should be created. Then the gap reveals the smallness of the overlap
of many--electron wave functions with-- and without such a  shift.
 The bosonization
 transformation \cite{Mat,Lut,Hal}
permits one to present the formation of the shift  as a creation of 
 plasmon modes in the system.   
The fact that the dispersion law of plasmons is linear 
in one dimension:  $\omega=sk$, results in a power--law behavior of 
 the density of states  $\nu(\omega)$. For
tunneling into the edge of a single--channel liquid 
the density of states has a form \cite{Kan,Gla}:
\begin{equation}
\label{sv}
\nu(\omega)\propto\omega^{\frac{s}{v_F}-1},  
\end{equation}
where $v_F$ is the Fermi velocity. Matveev an Glazman \cite{Gla} have traced
the evolution of the  exponent in (\ref{sv}) with increasing number of 
channels. 

Bosonization procedure can be   generically extended  to the
the  case of longitudinally inhomogeneous Luttinger liquid. 
If the spacial scale of inhomogeniety is much larger than the Fermi
wave length, $k_F^{-1}$,  it can be considered as a 
``scattering potential''
for the plasmon modes, so that a plasmon gets 
partially reflected  after propagation through
 the region of inhomogeniety.
If the boundaries of this   region are abrupt 
compared to the wave length of a
plasmon (but smooth on the scale of $k_F^{-1}$), the reflection
coefficient, as a function of frequency, exhibits the
 interference oscillations. Such oscillations were first considered in
Ref. \onlinecite{Fin}.
 In Ref. \onlinecite{Na} it was demonstrated
that the oscillations of the reflection coefficient
 lead to the oscillating structure in the 
density of states,  superimposed on the power--law increase.
    
In the present paper we consider another realization of inhomogeniety. 
Namely, we assume that the density of the Luttinger liquid is periodically
modulated with a period $a \gg k_F^{-1}$. We show that such a modulation
results in additional singularity in the tunnel density of states.
The origin of the singularity is a gap in the plasmon spectrum, which
opens due to the periodic modulation. 
In complete analogy with motion of an electron in a periodic potential,
the position of the center of the gap, $\omega_0$, is determined by the
condition $k(\omega_0)=\pi/a$, so that 
 $\omega_0 = \pi s/a$. We assume the modulation to be weak. Then the
width of the gap, $\Delta$, is much smaller than $\omega_0$, the
ratio $\Delta/\omega_0$ being
 proportional to the relative amplitude of the modulation. It is
also proportional to the interaction strength, since it is obvious that
without interaction the gap is absent. In the vicinity of $\omega_0$
the modulation--induced correction to the tunnel density of states,
$\delta\nu$, is
a function of $\tilde\omega=2(\omega -\omega_0)/\Delta$.
 For $|\omega-\omega_0|\gg \Delta$,  $\delta\nu(\tilde\omega)$
behaves  as a power law with the same exponent as in (\ref{sv}).
Since we study  the tunneling into the edge, the actual shape of     
$\delta\nu(\tilde\omega)$ appears to depend strongly
 on the {\em  phase} of
the modulation. This is because for plasmons with frequencies close to
$\omega_0$ the behavior of the field near the edge is determined by
this phase. The other reason is that,
with periodic modulation, the presence of an edge
leads to the formation of a localized plasmon mode with frequency 
within  the gap. For an electron  in a periodic potential this
fact was established more than 60 years ago \cite{Tam}.  The frequency
of the localized mode sweaps through the gap as the phase of modulation
changes. We show that the localized plasmon provides a contribution
to $\delta\nu(\tilde\omega)$ comparable to that from the extended modes.

The paper is organized as follows. In the next  section the form
of the Hamiltonian for an inhomogeneous Luttinger liquid is established.
 In Sect. III we derive a general
formula for inhomogeneity--induced  correction to the tunnel
density of states. 
In Sect. IV the
plasmon modes for the case of a weak sinusoidal modulation of the
electron concentration are found.  The results for $\delta\nu(\tilde
\omega)$ are analyzed in Sect. V.
 Section VI concludes the paper.

\section{Hamiltonian}
\label{sec:H}
In this section we will establish the form of the  Hamiltonian for
the Luttinger liquid with spatially varying concentration of
electrons $n_0(x)$.
We start from the classical equation of motion 
\begin{equation} 
\label{cl}
mn\frac{d^2u}{dt^2}=-en{\cal E}-\frac{dP}{dx},
\end{equation}
where $m$ is the electron mass; $u(x,t)$ is the displacement; 
 $n(x,t)$ is the electron density: $n(x,t)=n_0(x)+n_1(x,t)$, with $n_{1}$
describing the temporal  fluctuations; ${\cal E}(x,t)={\cal E}_0(x)+{\cal E}_1
(x,t)$ is the  actual electric field consisting of static and fluctuating
parts; $P=\pi^2\hbar^2n^3/3m$ is the hydrostatic pressure. The static
electric field can be eliminated from Eq. (1) using the equilibrium 
condition
\begin{equation} 
\label{eq}
e{\cal E}_0n_0=-\frac{dP}{dx}\Bigg |_{n=n_0(x)},
\end{equation}
which gives ${\cal E}_0(x)=-(\pi^2\hbar^2n_0/em) (dn_0/dx)$. Substituting
 this value into (1) and linearizing (1) with respect to $u$, ${\cal E}_1$
and $n_1$, we obtain
\begin{equation} 
\label{li}
mn_0\frac{d^2u}{dt^2}=-en_0{\cal E}_1 -\frac{\pi^2\hbar^2}{m}\left(
n_0^2\frac{dn_1}{dx} + n_0n_1\frac{dn_0}{dx}\right).
\end{equation}

Let $V(x)$ be the potential of electron--electron interactions, screened by 
the presence of a gate electrode. Then ${\cal E}_1$ can be presented as a
field created by the density fluctuations $n_1(x,t)$
\begin{equation}
\label{E}
e{\cal E}_1=-\frac{d}{dx}\int dx'V(x-x')n_1(x',t).
\end{equation}
As usually we will assume that the screening radius is much smaller than 
the characteristic spatial scale of the fluctuations. Then Eq. (\ref{E})
reduces to $e{\cal E}_1 = -V_0(dn_1(x,t)/dx)$, where $V_0=\int dx V(x)$.
    
As a next step $n_1$ is expressed through $u$ with the help of
 the continuity equation:
$n_1=-d(n_0u)/dx$. To derive the energy conservation law we multiply Eq. (1)
by $du/dt$. Then, using (\ref{E}), the first term in the right-hand side 
can be rewritten as
$-\frac{V_0}{2}\frac{d}{dt}\left[\frac{d(n_0u)}{dx}\right]^2+
V_0 \frac{d}{dx} \left[ \frac{d(n_0u)}{dx} \frac{d(n_0u)}{dt} \right]$.
The two remaining terms can be
combined into $\frac{\pi^2\hbar^2}{2m}\frac{d}{dt} 
\left[\frac{d}{dx}(n_0\frac {d(n_0u)^2}{dx}) - 
n_0(\frac{d(n_0u)}{dx})^2 \right]$. After bringing all these terms
into the left--hand side and integrating over $x$ from $0$ to $\infty$
(taking into account that $u(0,t)=0$), Eq. (1) reduces to $dE/dt = 0$, where
the energy $E$ is given by
\begin{equation}
\label{E=}
E=\int_{0}^{\infty}dx\biggl[\frac{mn_0(x)}{2}\biggl(\frac{du}{dt}\biggr)^2 + \frac{1}{2}
\biggl(V_0 + \frac{\pi^2\hbar^2}{m}n_0(x)\biggr)\biggl(\frac{d(n_0u)}{dx}\biggr)^2\biggr].
\end{equation}
The expression (\ref{E=})  for the energy allows one to write down the
corresponding Hamiltonian. Treating the displacement $u(x)$ as an operator
$\hat{u}(x)$ and
introducing the  conjugate momentum operator $\hat{p}(x)$, $[\hat{u}(x),\hat{p}
(x')]=i\hbar
\delta(x-x')]$, one gets
\begin{equation}
\label{H=}
\hat{\cal H}= \int_{0}^{\infty}dx\biggl[\frac{\hat{p}^2(x)}{2mn_0(x)} + \frac{1}{2}
\biggl(V_0 + \frac{\pi^2\hbar^2}{m}n_0(x)\biggr)\biggl(\frac{d(n_0\hat{u})}{dx}\biggr)^2\biggr].
\end{equation}     
If the concentration is constant, the last term in  the integrand takes the
standard form \cite{Gla} $\frac{1}{2}mn_0v_F^2(\frac{du}{dx})^2$, where 
$v_F=\pi n_0\hbar/m $ is the Fermi velocity. 

\section{General form of the correction to the density of states}
\label{sec:G}

In this section we will derive a general expression for the
correction, caused by the variation of the electron 
concentration $n_0(x)$, to the tunnel density of states at the edge of
the Luttinger liquid. The Hamiltonian (\ref{H=}) can be diagonalized by
the transformation
\begin{eqnarray}
\label{p=}
\hat{u}(x)= \sum_{\mu}\frac{1}{\sqrt{n_0(x)}}\Phi_{\mu}(x)\hat{Q}_{\mu}\\
\nonumber
\hat{p}(x)= \sum_{\mu}\sqrt{n_0(x)}\Phi_{\mu}(x)\hat{{\cal P}}_{\mu},
\end{eqnarray}   
where $\Phi_{\mu}(x)$ are the eigenfunctions of the equation
\begin{equation}
\label{Phi}
-\sqrt{n_0(x)}\frac{d}{dx}\biggl[\biggl(\frac{V_0}{m}+ 
\left(\frac{\pi\hbar}{m} \right)^2
n_0(x)\biggr)\frac{d}{dx} \left(\sqrt{n_0(x)}\Phi_{\mu} \right)\biggr]=
\Omega_{\mu}^2
\Phi_{\mu},
\end{equation}
and $\Omega_{\mu}^2$ are the eigenvalues. We assume that $\Phi_{\mu}$ are
normalized: 
$\int_0^{\infty} dx \Phi_{\mu}(x)\Phi_{\mu'}(x)=\delta_{\mu,\mu'}$, and
turn to zero at the edge: $\Phi_{\mu}(0)=0$. 
As a result of the transformation (\ref{p=}) the Hamiltonian (\ref{H=})
reduces to the system of harmonic oscillators 
\begin{equation}
\label{HH}
\hat{\cal H}= \sum_{\mu}\biggl[\frac{\hat{{\cal P}}_{\mu}^2}{2m}+
\frac{m\Omega_{\mu}^2}{2}\hat{Q}_{\mu}^2\biggr].                              
\end{equation}
with frequencies $\Omega_{\mu}$. Next, the transformation
(\ref{p=}) is applied to the operator $\Psi^{\dag}$ which creates an electron
at the edge $x=0$
\begin{equation}
\label{Psi}
\Psi^{\dag}=\exp\biggl(-\frac{i}{\hbar}\int_0^{\infty}dx\frac{\hat{p}(x)}{n_0(x)}\biggr)=\exp\biggl(-\sum_{\mu}\alpha_{\mu}\hat{{\cal P}}_{\mu}\biggr),
\end{equation}
where the coefficients  $\alpha_{\mu}$ are defined as
\begin{equation}
\label{alp}
\alpha_{\mu}=\int_{0}^{\infty}dx\frac{\Phi_{\mu}(x)}{\sqrt{n_0(x)}}.
\end{equation}
One can also express $\alpha_{\mu}$  explicitly through  
$(d\Phi_{\mu}/dx)|_{x=0}$. Dividing Eq. (\ref{Phi}) by $\sqrt{n_0(x)}$ and
integrating, one gets
\begin{equation}
\label{al=}
\alpha_{\mu}= -\frac{1}{\Omega_{\mu}^2}\left(\frac{V_0}{m}+
\left(\frac{\pi\hbar}{m} \right)^2
n_0(0)\right)\sqrt{n_0(0)}\left(\frac{d}{dx}\Phi_{\mu}\right)\Bigg |_{x=0}.
\end{equation}
With Hamiltonian and $\Psi^{\dag}$ having the form (\ref{HH}) and (\ref{Psi}),
the calculation of the density of states $\nu(\omega)$
 reduces to the standard procedure \cite{Mah} and results in
\begin{equation}
\label{nu}
\nu(\omega)=\frac{1}{\pi}{\mbox Re}\int_0^{\infty}dt e^{i\omega t}
<\Psi(t)\Psi^{\dag}(0)> = \frac{1}{\pi}{\mbox Re}\int_0^{\infty}dt e^{i\omega t}e^{-W(t)},
\end{equation}  
where the function $W(t)$ is the sum over eigenmodes
\begin{equation}
\label{W=}
W(t)=\sum_{\mu}\frac{m\alpha_{\mu}^2\Omega_{\mu}}{2\hbar}\biggl(1-e^{-i\Omega_{\mu}t}\biggr).
\end{equation} 

At this point we will make use of the assumption that the relative variation
of the concentration $n_0(x)$ is small: $n_0(x)=\bar{n}+\delta n(x)$;
$\delta n \ll \bar{n}$.
For the average concentration, $\bar{n}$,
 the eigenfunctions of Eq. (\ref{Phi}),
which satisfy the boundary condition, are $\propto \sin kx$, 
the corresponding eigenvalues being $s^2k^2$, where 
\begin{equation}
\label{s=}
s=v_{F}\sqrt{1+\frac{V_0\bar{n}}{mv_F^2}}
\end{equation}
is the plasmon velocity. Then the function $W(t)$ takes the standard
form
\begin{equation}
\label{WW}
W_0(t)=\frac{s}{v_F}\int_0^{\infty}\frac{dk}{k}
(1-e^{-iskt})e^{-r_0k}=
\frac{s}{v_F}\ln\biggl(\frac{r_0 + ist}{r_0}\biggr),
\end{equation}
where, as usually, the cutoff parameter $e^{-r_0k}$ is introduced to
insure the convergence of the integral at $k\rightarrow \infty$. With 
$W_0(t)$ given by (\ref{WW}),
 Eq. (\ref{nu}) reproduces the result (\ref{sv}).
 Treating the difference $W(t)-W_0(t)$ as a perturbation
and expanding the exponent in (\ref{nu}) to the first order, one gets the
following correction to the density of states
\begin{equation}
\label{dnu}
\delta\nu = -\frac{1}{\pi}{\mbox Re}\int_0^{\infty}dt e^{i\omega t}
\biggl(\frac{r_0}{r_0+ist}\biggr)^{s/v_F}\biggl[W(t)-W_0(t)\biggr].
\end{equation}
This expression can be simplified if one uses the identity
\begin{equation}
\label{Ga}
(r_0 + ist)^{s/v_F}=\frac{1}{\Gamma(s/v_F)}\int_0^{\infty}dzz^{\frac{s}
{v_F}-1}e^{-z(r_0 + ist)}.
\end{equation}   
Substituting (\ref{Ga}) into (\ref{dnu}) and performing integration
first over $t$ and then over $z$, one gets
\begin{equation}
\label{main}
\delta\nu = \frac{1}{\Gamma(s/v_F)}\biggl(\frac{r_0}{s}\biggr)^{s/v_F}
\Biggl[\sum_{\mu}\frac{m\alpha_{\mu}^2\Omega_{\mu}}{2\hbar}(\omega-\Omega_{\mu})^
{\frac{s}{v_F}-1} - \frac{s}{v_F}\int_0^{\omega/s}\frac{dk}{k}(\omega-sk)^
{\frac{s}{v_F}-1} \Biggr].
\end{equation}  
The divergence of the integral at small $k$ is canceled by the similar
divergence in the first term.
Further calculations require the knowledge of the eigenfrequencies
$\Omega_{\mu}$. For weak periodic modulation they are found in
the next Section.

\section{Plasmon spectrum}
\label{sec:P}
To find the eigenvalues of Eq. (\ref{Phi}) it is convenient to 
introduce a function
\begin{equation}
\label{tilda}
\tilde\Phi_{\mu} = \Phi_{\mu}\sqrt{n_0(x)(n_0(x)+V_0m/\pi^2\hbar^2)}. 
\end{equation} 
Then the equation for $\tilde\Phi_{\mu}$ takes the form
\begin{equation}
\label{dif}
\frac{d^2\tilde\Phi_{\mu}}{dx^2}=\Biggl[\frac{1}{4}\biggl(\frac{dn_0/dx}{n_0(x)+V_0m/\pi^2\hbar^2}\biggr)^2 - \frac{n_0(x)}{(n_0(x)+V_0m/\pi^2\hbar^2)}
\Biggl( \biggl(\frac{\Omega_{\mu}m}
{\pi\hbar n_0(x)}\biggr)^2 +\frac{d^2n_0/dx^2}{2n_0(x)}\Biggr) \Biggr]
\tilde\Phi_{\mu}.
\end{equation}
Let us specify the modulation profile:
 $\delta n(x)= -n_{M}\cos(\sigma x + \varphi)$, where
$\varphi$ is the phase and $\sigma = 2\pi/a$.
For such a modulation and $n_M \sim \bar{n}$ the plasmon spectrum was
studied numerically in Ref. \onlinecite{Men}. If $n_M \ll \bar{n}$
the analytical solution can be obtained. After expanding the ``potential
energy'' in (\ref{dif}) to the first order in $n_M/\bar{n}$, the
equation reduces to the conventional Mathieu equation\cite{A.E}
\begin{equation}
\label{Om}
\frac{d^2\tilde\Phi_{\mu}}{dx^2} + \biggl[\frac{\sigma^2\theta}{2}\cos
(\sigma x + \varphi) + \biggl(\frac{\Omega_{\mu}}{s}\biggr)^2\biggr]
\tilde\Phi_{\mu}=0,
\end{equation}
where the dimensionless  parameter $\theta$ is defined as
\begin{equation}
\label{th}
\theta=\frac{2n_M}{\bar{n}}\biggl[\biggl(\frac{\Omega_{\mu}}{\sigma s}\biggr)^2\left(\frac{v_F^2}{s^2}+1 \right)-\frac{v_F^2}{2s^2}\biggr].
\end{equation}
Since $\theta$ is proportional to the relative modulation, we have
$\theta \ll 1$, so that the gap in the spectrum of eigenfrequencies
$\Omega_{\mu}$ is narrow.  The center of the gap is
determined by the condition $\Omega_{\mu}/s = \sigma/2$. This allows
to simplify the expression for $\theta$ by the substitution
$\Omega_{\mu}=\sigma s/2$. Then (\ref{th}) takes the form
\begin{equation}
\label{the}
\theta=\frac{n_M}{2\bar{n}}\biggl(1-\frac{v_F^2}{s^2}\biggr)=\biggl(\frac{n_M}{2\bar{n}}\biggr)
\frac{V_0m/\pi^2\hbar^2}{\bar{n}+V_0m/\pi^2\hbar^2}.
\end{equation}  
We emphasize again that $\theta$ is proportional to the interaction
strength $V_0$; this  supports the obvious observation
 that there is no gap for non--interacting electrons.
Note also that, in principle, there are higher--order gaps
 in the spectrum of Eq. (\ref{Om}). They are centered at 
$\Omega_{\mu}=ps\sigma/2$ ($p=2,3...$) with the widths proportional
to $(n_M/\bar{n})^p$. However, they do not describe the higher--order
gaps in the spectrum of Eq. (\ref{dif}), since (\ref{Om}) was derived
from (\ref{dif}) using the first--order expansion.  

As it was mentioned in the Introduction, the eigenmodes of Eq. (\ref{Om}),
satisfying the condition $\tilde\Phi_{\mu}(0)=0$,
 can be of two types: localized and extended.  

\subsection{Localized mode} 
\label{sec:L}

Substituting into (\ref{Om})
\begin{equation}
\label{loc}
\tilde\Phi=2\bar{n}\frac{s}{v_F} \sqrt{\gamma}e^{-\gamma x}\sin\frac{
\sigma x}{2},
\end{equation}
where the prefactor insures the normalization, and neglecting the
``non--resonant'' terms $\propto \sin(\frac{3\sigma x}{2} + \varphi)$,
we get the following expressions for the frequency and the decay 
constant of a localized mode
\begin{eqnarray}
\label{fr}
\Omega &=& \frac{s\sigma}{2}\sqrt{1+\theta\cos\varphi}\approx \frac{s\sigma}{2}\left(1+\frac{\theta}{2}\cos\varphi \right)\\ \nonumber
\gamma &=& -\frac{\sigma\theta}{4}\sin\varphi.
\end{eqnarray}
We see that the localized mode exists only if $\sin\varphi < 0$, when
$\gamma$ is positive.
As $\varphi$ changes within the interval $[\pi,2\pi]$, the frequency $\Omega$ moves from the lower to the upper boundary of the gap. The decay
constant  of the localized mode turns to zero as its
 frequency merges with the continuum.
 To calculate the contribution of the localized 
mode to $\delta\nu$, we need the constant $\alpha$. Substituting 
(\ref{fr}) into (\ref{al=}), we obtain  
\begin{equation}
\label{pha}
\alpha=-\frac{4}{\sigma}\sqrt{\frac{\gamma}{\bar{n}}}.
\end{equation}

\subsection{Extended modes}
\label{sec:E}

Let $L$ be the normalization length. We search for solution of (\ref{Om})
in the form 
\begin{equation}
\label{ext}
\tilde\Phi_{\mu}=\bar{n}\frac{s}{v_F}\sqrt{\frac{2}{L(\beta_1^2 +
\beta_2^2)}}\Biggl[\beta_1\cos\biggl((\frac{\sigma}{2} + \kappa_{\mu})x +
\psi_1\biggr)+ \beta_2\cos\biggl((\frac{\sigma}{2} - \kappa_{\mu})x +
\psi_2\biggr) \Biggr].
\end{equation}
The boundary condition requires that
\begin{equation}
\label{boundary}
\beta_1\cos\psi_1 +\beta_2\cos\psi_2=0.
\end{equation}
Substituting (\ref{ext}) into (\ref{Om}),
 we establish the following relation between  phases
\begin{equation}
\label{rel}
\psi_1 +\psi_2=\varphi.
\end{equation}
The system of coupled equations for coefficients $\beta_1$, $\beta_2$
takes the form
\begin{eqnarray}
\label{sy}
\Biggl[\biggl(\frac{\Omega_{\mu}}{s}\biggr)^2-\biggl(\frac{\sigma}{2}+
\kappa_{\mu}\biggr)^2\Biggr]\beta_1 + \frac{\sigma^2\theta}{4}\beta_2 &=& 0
\nonumber \\
\Biggl[\biggl(\frac{\Omega_{\mu}}{s}\biggr)^2-\biggl(\frac{\sigma}{2}-
\kappa_{\mu}\biggr)^2\Biggr]\beta_2 + \frac{\sigma^2\theta}{4}\beta_1 &=& 0.
\end{eqnarray} 
Solution of the system yields the plasmon spectrum
\begin{equation}
\label{spe}
\Omega_{\mu}=\omega_0 + \lambda \sqrt{\kappa_{\mu}^2 s^2 + \frac{\Delta^2}
{4}},
\end{equation}
where $\omega_0=s\sigma/2$, and the width of the gap, $\Delta$,
is given by
\begin{equation}
\label{Dl}
\Delta=\frac{s\sigma\theta}{2}=\theta\omega_0,
\end{equation}
so that the smallness of  $\theta$ guarantees
 that  $\Delta/\omega_0 \ll 1$. Parameter $\lambda$ takes the values
$\pm 1$ depending on the sign of $\kappa_{\mu}$
\begin{equation}
\label{la}
\lambda= {\mbox sign} (\kappa_{\mu}).
\end{equation}
The plasmon spectrum (\ref{spe}) is shown in Fig. 1.  
The  phase of modulation $\varphi$  does not affect the spectrum,
but it reveals itself in the parameter $\alpha_{\mu}$, which for
$\tilde\Phi_{\mu}$, given by (\ref{ext}), takes the form
\begin{equation}
\label{all}
\alpha_{\mu}=\frac{2}{\sigma}\sqrt{\frac{2}{L\bar{n}}}\frac{\beta_1\cos
\psi_1 + \beta_2\cos\psi_2}{(\beta_1^2 + \beta_2^2)^{1/2}}.
\end{equation}    
The last factor in (\ref{all}) can be expressed through $\varphi$
using the relations (\ref{boundary}) and (\ref{rel})
\begin{equation}
\label{expr}
\frac{\beta_1\cos
\psi_1 + \beta_2\cos\psi_2}{(\beta_1^2 + \beta_2^2)^{1/2}}=
\frac{\beta_1^2 -\beta_2^2}{(\beta_1^2+\beta_2^2)^{1/2}(\beta_1^2+\beta_2^2
+2\beta_1\beta_2\cos\varphi)^{1/2}}.
\end{equation}
Expressing the ratio $\beta_2/\beta_1$ with the use of
 Eqs. (\ref{sy}), (\ref{spe}),
we finally obtain
\begin{equation}
\label{fin}
\frac{\beta_1\cos
\psi_1 + \beta_2\cos\psi_2}{(\beta_1^2 + \beta_2^2)^{1/2}}=-\frac{s\kappa_{\mu}}
{\sqrt{s^2\kappa_{\mu}^2+\frac{\Delta^2}{4}\sin^2\varphi}}
\Biggl[1+\lambda\cos\varphi\frac{\Delta /2}{\sqrt{s^2\kappa_{\mu}^2+\frac{\Delta^2}{4}}}\Biggr]^{1/2}.
\end{equation}
Discreteness of $\kappa_{\mu}$ is related to the final normalization length.
Summation over $\mu$ reduces to integration with the help of the relation
$\delta\kappa_{\mu} = \pi/L$.

\section{Asymptotes and numerical results}
\label{sec:A}

The resulting expression for $\delta\nu$ emergies after substitution
of (\ref{fr}), (\ref{pha}), (\ref{spe}) and (\ref{all})
 into Eq. (\ref{main}). It is convenient to present this expression
by introducing the relative deviation of the frequency $\omega$ from
the center of the gap
\begin{equation}
\label{dev}
\tilde\omega = \frac{2(\omega - \omega_0)}{\Delta}.
\end{equation}
Then we obtain
\begin{equation}
\label{final}
\delta\nu (\tilde\omega)=\frac{1}{\Gamma (s/v_F)}\biggl(\frac {r_0 \theta}
{s}\biggr)^{s/v_F}\left(\frac{s}{v_F}\right)\omega_0^{\frac{s}{v_F}-1}F(\tilde\omega),
\end{equation}
where the dimensionless function $F(\tilde\omega)$ has the form
\begin{equation}
\label{F}
F(\tilde\omega)=\left[-\pi\sin\varphi(\tilde\omega - \cos\varphi)^{\frac
{s}{v_F}-1} - \frac{1}{2}f_1(\tilde\omega)+ \frac{\cos\varphi}{2}f_2(\tilde
\omega)\right],
\end{equation}
and the functions $f_1$ and $f_2$ are defined as
\begin{eqnarray}
\label{f1}
f_1(\tilde\omega) &=& \int_{-\infty}^{\infty}dv\left[(\tilde\omega - v)^{\frac{s}{v_F}-1} - \frac{v^2}{v^2 + \sin^2\varphi}(\tilde\omega - \lambda_{v}
\sqrt{v^2 + 1})^{\frac{s}{v_F}-1}\right],\\
\label{f2}
f_2(\tilde\omega) &=& \int_{-\infty}^{\infty}dv\frac{\lambda_{v}v^2}{(v^2 +
\sin^2\varphi)\sqrt{v^2 + 1}}\left(\tilde\omega - \lambda_{v}\sqrt{v^2 +1}
\right)^{\frac{s}{v_F}-1}.
\end{eqnarray}
Similar to (\ref{la}), in (\ref{f1}), (\ref{f2})  $\lambda_{v}= sign(v)$;
the form $(...)^{\frac{s}{v_F}-1}$ is defined only when the argument is positive and it is zero otherwise. 

We  remind that the first term in $F(\tilde\omega)$ is present only if
$\sin\varphi < 0$. This term desribes a power--law anomaly in the
density of states originating from the localized plasmon. The second and 
the third terms represent the contribution from the continuous
plasmon  spectrum outside  the gap. 

In the integral $f_1$ two terms
of the integrand almost cancel each other  at large negative $v$, indicating
that the modulation--induced correction to the spectrum is important only
in the region of the gap.
One can see that the integral
$f_2$ divergies as $v\rightarrow - \infty$, since for large negative $v$ 
the integrand behaves as $-(-v)^{\frac{s}{v_F}-2}$.
The origin of the divergency is that in (\ref{f1}), (\ref{f2})
 we have used the
expansion of the spectrum in the vicinity of the gap: $|k-\frac{\sigma}{2}
| \ll \sigma$, so the integration should be cut at  $v \sim -  1/\theta$.
 Note however, that the divergent part is frequency--independent and
causes only a small correction to the background density of states $\nu
(\omega_0)$. To study the frequency dependence one should subtract
this divergent part by considering the difference $\tilde f_2 = f_2(\tilde
\omega)-f_2(0)$, which converges. 

For large  positive $\tilde\omega$ (outside the gap)
 all three contributions to $F(\tilde\omega)$
behave as $\tilde\omega^{\frac{s}{v_F}-1}$ with prefactors depending on
the phase of modulation $\varphi$. For integrals $f_{1}$ and $f_{2}$ the 
following asymptotic behavor in the limit $\tilde\omega \gg 1$ can
be derived   
\begin{eqnarray}
\label{deri}
f_1 &\approx& \pi|\sin\varphi|  \tilde\omega^{\frac{s}{v_F}-1}, \nonumber \\
\tilde f_2 &=& f_2(\tilde\omega)-f_2(0)\approx \frac{\pi} {\tan(\frac{\pi s}{v_F})}
\tilde\omega^{\frac{s}{v_F}-1}.
\end{eqnarray}
In the opposite limit $\tilde\omega < 0$, $|\tilde\omega| \gg 1$ the
contribution from the localized plasmon is absent and the
 integral $f_1$ falls off as $|\tilde\omega|^{\frac{s}{v_F}-2}$. The only
contribution to $F(\tilde\omega)$ in this limit comes from $\tilde f_2$
which has the following asymptotics
\begin{equation}
\label{asy}
\tilde f_2 \approx -\frac{\pi}{\sin(\frac{\pi s}{v_F})}|\tilde\omega|^{\frac 
{s}{v_F}-1}. 
\end{equation}  

The frequency dependence of the correction $\delta\nu(\tilde\omega)$,
 calculated numerically from Eq. (\ref{F}) for 
various values of the phase $\varphi$,
 is shown in Fig. 2. It can be seen that some  of the  curves have cusps
at frequencies,   corresponding to the boundaries of the gap $\tilde
\omega = \pm 1$. The cusps in the curves $\varphi = 11\pi/8 $
and $\varphi = 3\pi/2$ have their origin  in the localized   plasmon.

\section{Conclusion}
\label{C}
The singularity in the tunnel density of states,
 caused by the periodic  modulation of the electron concentration,
should reveal itself as an anomaly in the differential resistance
$dI/dV$ at voltage $eV =\hbar\omega_0 = \pi\hbar s/a$, which  depends on 
the period of modulation and the plasmon velocity. We considered the
case of a weak sinusoidal modulation. If higher harmonics with periods
$a/p$ are present
in $n_0(x)$, the  gaps in the plasmon spectrum at $\omega = p\omega_0$
 (and correspoding anomalies in $dI/dV$
at $eV = \pi p\hbar s/a$) should emerge. 

The strength of the anomaly is governed by the dimensionless 
parameter $\theta$ (\ref{th}), which  we assumed to be small. To sense the
magnitude of the anomaly it is convenient to relate $\delta\nu$ to
the   density of states at  $\omega = \omega_0$ in the absence of
modulation: $\delta\nu/\nu(\omega_0)=\theta^{s/v_F}F(\tilde\omega)$.

In our calculations we assumed the interactions to be short--range.
This implies that the interaction radius $\sim r_0$ should
be much smaller than the period of modulation $a$.

As a final remark, we considered a single--channel Luttinger
liquid. Increasing of the number of channels would result in the
new anomalies in the tunnel density of states, caused by
resonances within each channel. Besides,  modulation would
couple the plasmons from different channels, thus causing additional
anomalies.

\acknowledgments

The authors are grateful to L. I. Glazman for very useful discussions.
We also appreciate the discussions with D. L. Maslov and A. A. Odintsov.

\begin{figure}
\caption{Plasmon spectrum of a periodically modulated Luttinger liquid.}
\label{fig1}
\end{figure}
\begin{figure}
\caption{The normalized correction to the density of states, 
$\delta\nu/\nu(\omega_0)\theta^{s/v_F}$, is plotted as a function of the
dimensionless frequency $\tilde\omega$ for $\frac{s}{v_F}=1.2$
and  different values of the phase of modulation $\varphi$. (a) $\varphi =0$  
(solid curve), $\varphi =\pi/4$ (long--dashed curve),  $\varphi =\pi/2$
 (dash--dotted curve). (b)  $\varphi =3\pi/4$  
(solid curve), $\varphi =\pi$ (long--dashed curve),  $\varphi = 11\pi/8$
 (dash--dotted curve).
(c) $\varphi =3\pi/2$  
(solid curve), $\varphi =7\pi/4$ (long--dashed curve), 
 $\varphi =15\pi/8$ (dash--dotted curve).}
\label{fig2}
\end{figure}
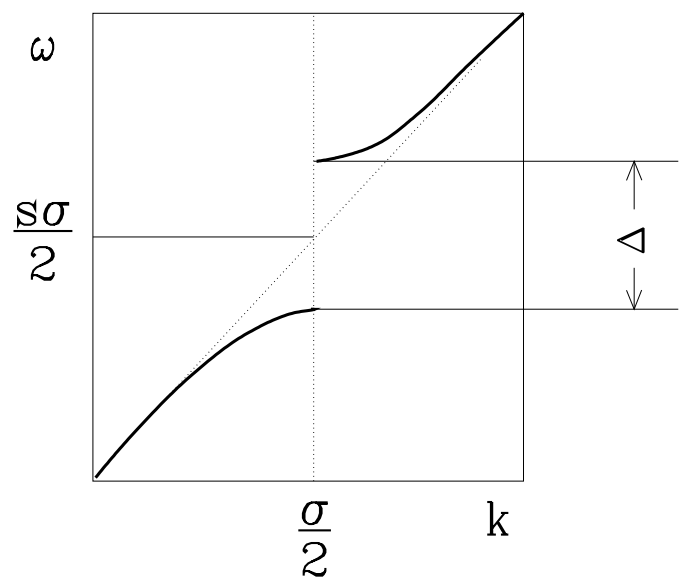
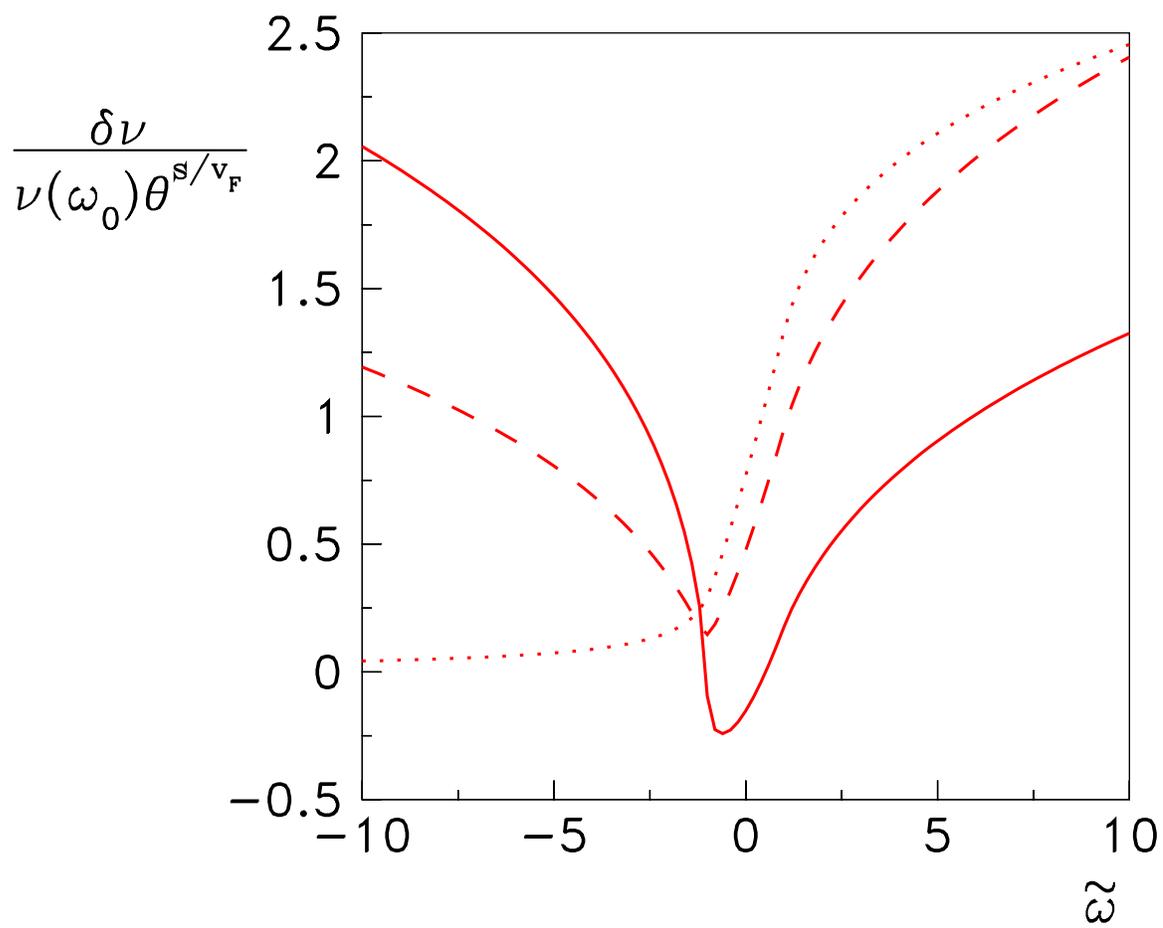
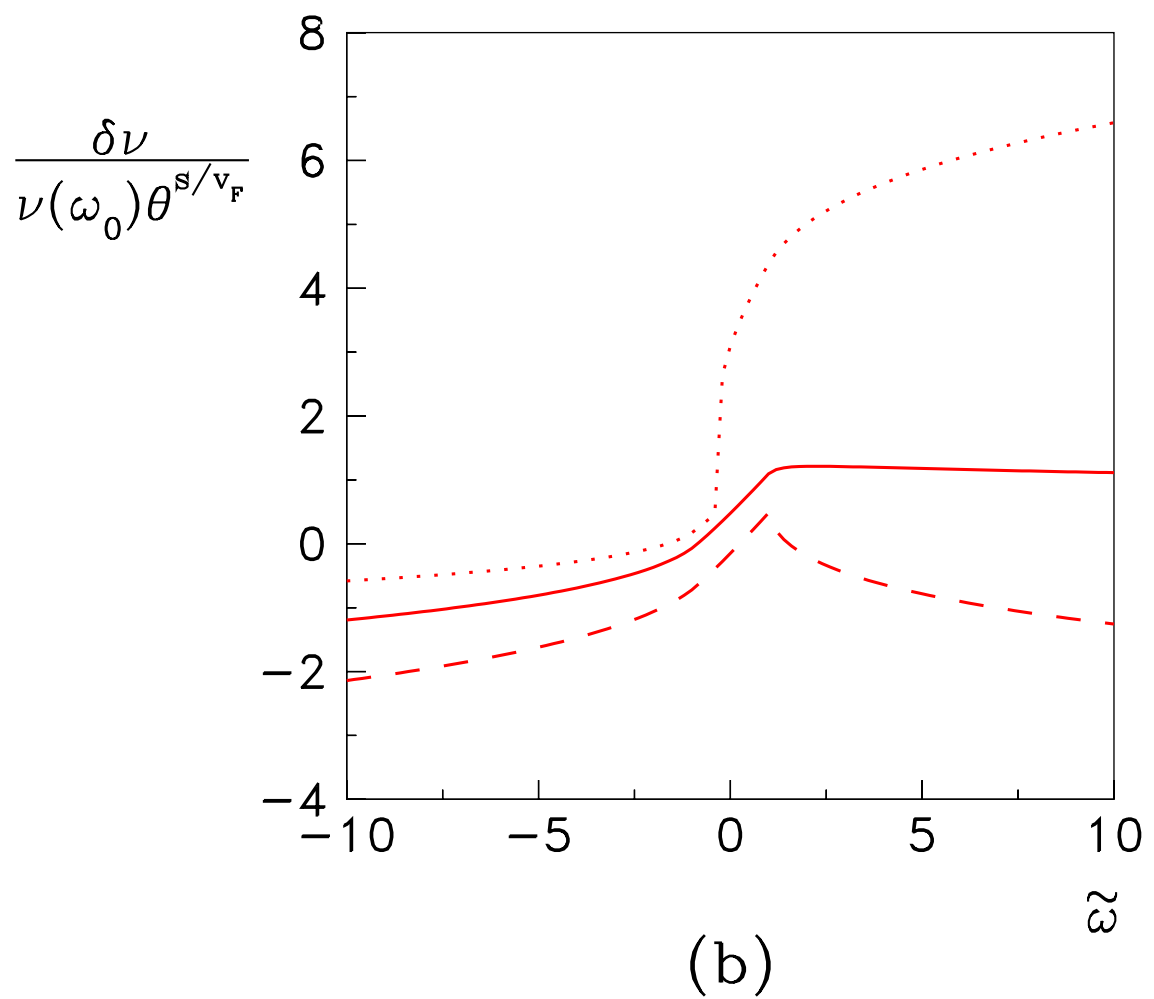
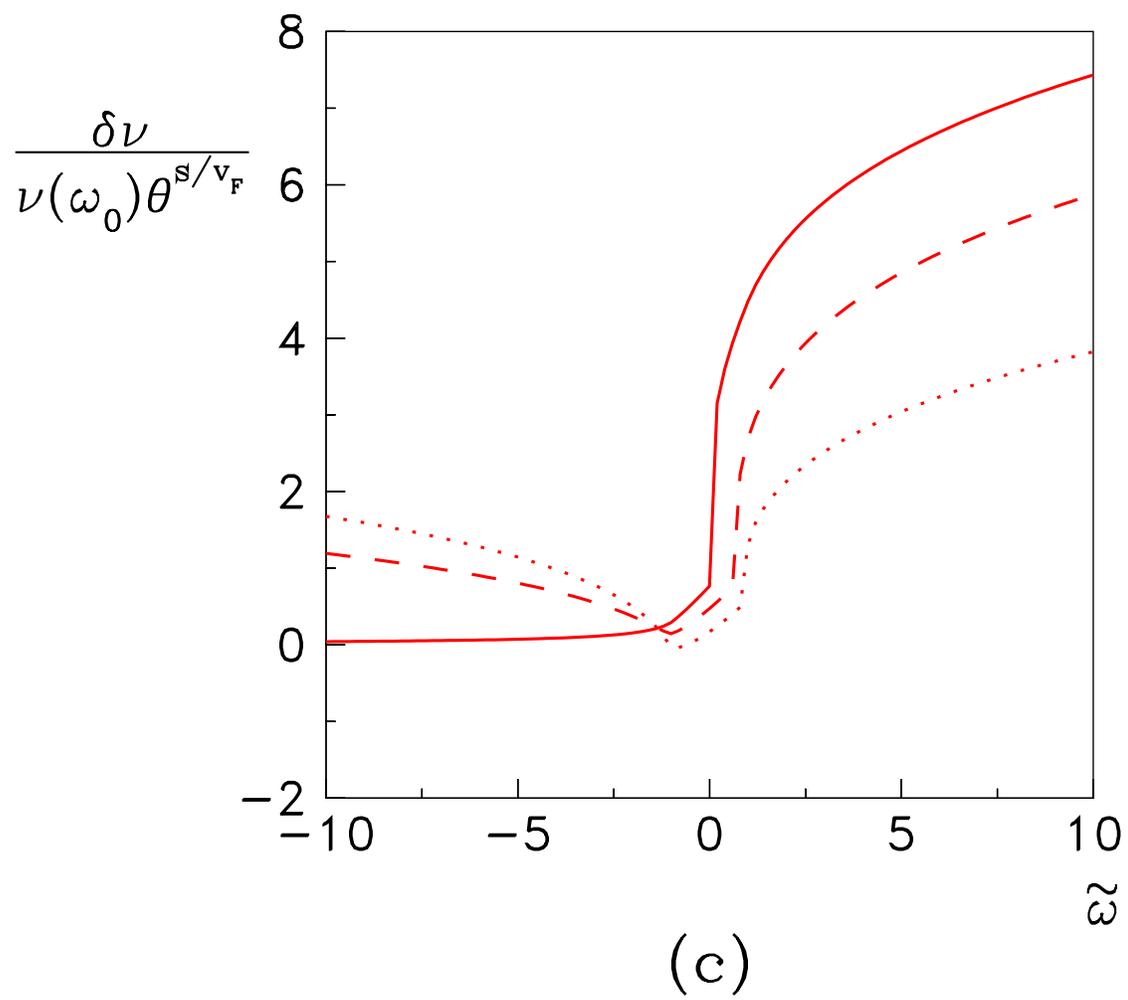
\end{document}